\documentclass[twocolumn]{aastex63}

\usepackage[version=4]{mhchem}

\usepackage[T1]{fontenc}

\shorttitle{Inner Disk Water Vapor Enrichment from Pebble Delivery}
\shortauthors{Easterwood et al.}

\begin{document}

\title{Water Enrichment from Pebble Drift in Disks with Gap-forming Planets}

\correspondingauthor{Whittney Easterwood}
\email{weasterwood@uidaho.edu}

\author[0009-0002-1445-9931]{Whittney Easterwood}
\affiliation{Department of Physics \\
Texas State University \\
749 N Comanche St, San Marcos, TX 78666, USA}

\author[0000-0002-5067-1641]{Anusha Kalyaan}
\affiliation{Department of Physics \\
Texas State University \\
749 N Comanche St, San Marcos, TX 78666, USA}

\author[0000-0003-4335-0900]{Andrea Banzatti}
\affiliation{Department of Physics \\
Texas State University \\
749 N Comanche St, San Marcos, TX 78666, USA}

\newcommand{\anusha}[1]{\textcolor{magenta}{\textbf{Anusha:} #1}}
\newcommand{\whittney}[1]{\textcolor{violet}{\textbf{Whittney:} #1}}
\newcommand{\ab}[1]{\textcolor{orange}{\textbf{Andrea:} #1}}
\newcommand{\bibtex}{\textsc{Bib}\!\TeX} %

\begin{abstract}

Volatiles like \ce{H2O} are present as ice in solids in the outer cold regions of protoplanetary disks and as vapor in the warm inner regions within the water snow line. Icy pebbles drifting inwards from the outer disk sublimate after crossing the snow line, enriching the inner disk with solid mass and water vapor. Meanwhile, proto-planets forming within the disk open gaps in the disk gas, creating traps against the inward drift of pebbles and in turn reducing water enrichment in the inner disk. Recent disk observations from millimeter interferometry and infrared spectroscopy have supported this broad picture by finding a correlation between the outer radial distribution of pebbles and the properties of inner water vapor spectra. In this work, we aim at further informing previous and future observations by building on previous models to explore pebble drift in disks with multiple gaps. We systematically explore multiple gap locations and their depths (equivalent to specific masses of planets forming within), and different particle sizes to study their impact on inner disk water enrichment. We find that the presence of close-in deep gaps carved by a Jupiter-mass planet is likely crucial for blocking icy pebble delivery into the inner disk, while planets with lower masses only provide leaky traps. We also find that disks with multiple gaps show lower vapor enrichment in the inner disk. Altogether, these model results support the idea that inner disk water delivery and planet formation are regulated by the mass and location of the most massive planets.
\end{abstract}

\section{Introduction}
The discovery of thousands of exoplanets in the last decade has revolutionized the field of planet formation. These planets as well as ones within our solar system all formed within protoplanetary disks, i.e., disks of gas and dust surrounding a young forming star. Observations of many protoplanetary disks using millimeter interferometry show substructures such as gaps and cavities \citep[][]{huang18a,long18,andrews2020, vandermarel2023}
that may have been carved by planets \citep[][]{paarmel2006, rice2006,zhu2012,bae2023review}, while a few disks have been discovered to harbor young planets within these structures, revealing ongoing planet formation \citep{Keppler18,haffert2019,isella2019,benisty21}.

Gaps created by the first young planets can affect the subsequent radial transport of solid material in the disk. Once growing planets are sufficiently massive to create a gap in the disk gas, the change in the surrounding pressure gradient stops the inward drift of pebbles \citep[][]{paarmel2006, morbinesv12, ataiee18, bitsch18a}. Pebbles are trapped beyond the gap and unable to complete their passage into the inner disk, where ice in them can sublimate to vapor. Several works have studied the link between delivery of icy pebbles from the outer disk, and inner disk chemistry and vapor enrichment  \citep{cuzzizahnle2004,cieslacuzzi2006,piso15,Booth2017,bitsch2021,kalyaan21,schneiderbitsch21, kalyaan23,mah2023}. Observationally, tracking water through icy pebbles to vapor in the inner regions requires combining observations sensitive to different regions in protoplanetary disks. Millimeter interferometry (e.g., with the Atacama Large Millimeter Array or ALMA) is especially sensitive to structure such as gaps and rings in the outer disk ($>$ 10 au). Complementarily, infrared (IR) spectroscopy provides a window into the gaseous chemistry of the warm inner disk where several volatiles such as \ce{H2O}, CO and \ce{CO2} are abundant \citep[e.g.][]{pont2014}. Since the presence of pebble traps in disks can prevent rapid inward pebble drift and keep large disks large \citep[e.g.][]{pinilla12,dullemond18,rosotti19,appelgren20,zormpas22}, the outer millimeter dust disk radius can be used as a proxy for the efficiency of pebble drift in a given disk, under the assumption that the outermost gap sets the disk radius, and that it is deep enough to effectively trap dust beyond it. The discovery of an anti-correlation between infrared water luminosity (corrected for accretion) and millimeter dust disk radii was interpreted in previous work as evidence for compact disks to have higher water column densities than extended disks \citep{banzatti2020}. More recent higher-resolution mid-IR spectra from the James Webb Space Telescope (JWST) showed that indeed compact disks have a cool water excess near the water snow line compared to large structured disks, which is consistent with water vapor enrichment by sublimation of icy pebbles drifting from the outer disk \citep{banzatti2023}.

\citet{kalyaan21} carried out a detailed modeling study to investigate how the delivery of ices into the inner disk brought in by drifting icy pebbles could enrich the inner disk with vapor, and how this enrichment was affected by the presence of gaps. They found that the inner disk vapor enrichment was a time-evolving quantity, and was sensitive to the radial location of the gap, and the size of the dust particles that were drifting in. They also found a `sweet spot' or a unique radial location for each particle size where they would be most efficiently trapped beyond the gap located there. They argued that these regions led to the least vapor enrichment in the inner disk as closer-in gaps (of same depth) would be more leaky, while gaps farther out would trap much less ice material beyond them. \citet{kalyaan21}, however only performed a limited parameter study with their model to test the effect of the presence of a single gap on inner disk vapor enrichment, and did not consider multiple gaps in disks that are routinely observed by ALMA \citep[e.g.,][]{andrews2020}. Moreover, they also only studied the effect of a single gap depth and did not explore gaps that could be shallower or deeper. In this work, we build on \citet{kalyaan21} by using the same disk evolution model to explore the effect of both multiple gaps and gaps with different depths, on the delivery of pebbles of different sizes, where we assume that different masses of forming planets produce gaps with different depths. Finally, we study the water vapor enrichment in inner regions of disks that have gaps of varying depths, corresponding to possible configurations of early planetary systems.

Our paper is organized as follows: Section \ref{sec:Methods} briefly describes our disk evolution model as well as the design of our parameter study; Section \ref{sec:results} describes the results of our simulations. Section \ref{sec:discussion} first describes the key insights that can be gained from our work, and then explores vapor enrichment in disks with gap-forming planets. Finally, Section \ref{sec:conclusions} summarizes the conclusions of our study.

\section{Methods}\label{sec:Methods}

In this work, our goal is to study the effect of multiple gaps on water vapor enrichment in the inner regions of protoplanetary disks. In this section, we first summarize the key features of our model and then discuss in detail specific aspects relevant to this work.

\begin{figure*}
	\centering
	\includegraphics[scale=.34]{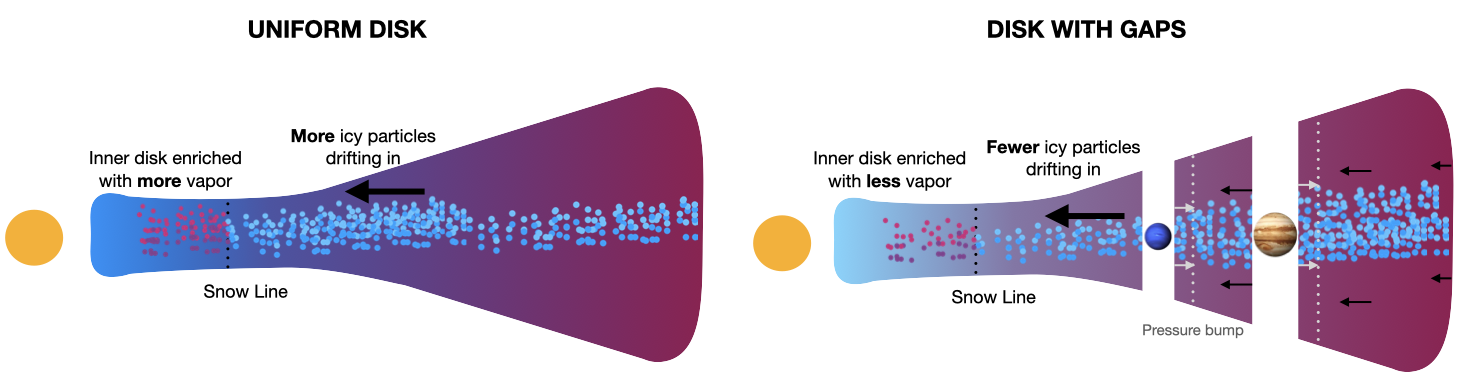}
	\caption{A schematic figure of our physical model with the transport of particles and vapor in a uniform disk (\textit{left}) and a disk with gaps created by young forming planets (\textit{right}).}
	\label{fig:1}
\end{figure*}

\subsection{Numerical Model}
Our numerical model is described in detail in \citet{kalyaan21}. We use a 1-D standard $\alpha$-disk evolution model, with an initial outer disk radius of 500 au and a characteristic radius of 70 au \footnote{Most of the disk mass is concentrated within 70 au, beyond which the radial profile of the surface density tapers exponentially}, evolved over several million years.
Over this underlying gaseous disk, we include the radial transport of a uniformly sized ice-bearing dust particle population, including their drift, diffusion, and advection (based on \citet[]{desch2017} and similar to \citet[]{birnstiel2010}).  We also use the same treatment for vapor transport via advection and diffusion as used in \citet{kalyaan21}, specified in detail in \citet[][see Equation 48]{desch2017} as well as the same treatment for condensation and evaporation of the ice within these particles at the snow line as used in previous works \citep[][]{kalyaan21,kalyaan23}. As in \citet{kalyaan21}, we neglect accretion heating, and use the following radial temperature profile: 
\begin{equation}\label{tempeqn2}
T(r) = 118 \, \left( \frac{r}{ 1 \, {\rm AU}} \right)^{-0.5} \, {\rm K},
\end{equation}

Our water snow line is therefore roughly at 0.3 au.

The model incorporates structure in the form of gaps in the gaseous disk that create pressure bumps trapping dust particles beyond the gaps. Overall, our model is able to track the vapor enrichment in the inner disk, by modeling the passage of icy dust particles as they drift inwards into the inner disk and lose ice to sublimation within the water snow line. If dust particles encounter a gap, then they are trapped in the pressure bump beyond the gap, unable to drift into the inner disk and generate water vapor there. \citet{kalyaan21} investigated how radial transport of particles of different sizes was affected by the presence of a single gap at different radial locations, in turn affecting the water vapor content in the inner disk. Here, we expand this previous work by investigating the effects of multiple gaps on inner disk vapor enrichment, as well as the effect of the depth of each gap.  Figure \ref{fig:1} depicts the main features of our model in a radial slice of a disk including the trapping of icy pebbles in pressure bumps beyond gaps carved by different planets and sublimation of ice to vapor at the water snow line. The model begins without any structure, with unhindered pebble drift proceeding with time. At 0.1 Myr, gaps are instantaneously formed in the disk, and dust particles (earlier allowed to evolve freely) begin to get trapped beyond the gaps. All our gaps are sufficiently outside of the water snow line.
Finally, our simulations are evolved for a period of 6 Myr. 

\subsection{Exploring Multiple Gaps}

To build on the previous work in \citet{kalyaan21}, we include multiple gaps at different radial locations in the disk model. To incorporate gaps in the disk gas, we follow the same treatment as in several previous works \citep{dullemond18,desch18,stamm19,kalyaan21,kalyaan23} for creating gaps in the disk gas with the help of Gaussian peaks in the turbulent viscosity $\alpha$ profile for each gap (see Figure \ref{fig:2}) as follows:

\begin{equation}
    \alpha(r) = \alpha_{\rm 0} + (\alpha_{\rm gap} - \alpha_{\rm 0})\, \exp\,(-x^2) \,.
\end{equation}  

Here, $x = (r - r_{\rm gap})$ / gap width, where $r$ is the distance from the star, r$_{\rm gap}$ is the radial location of the gap, and gap width is assumed to be equivalent to 2 $\times$ H. H is the scaleheight given as: H = c$_{\rm s}$/$\Omega$, where c$_{\rm s}$ is the local sound speed and $\Omega$ is the Keplerian angular frequency. $\alpha_{\rm 0}$ = 1 $\times$ 10$^{-3}$ is the constant $\alpha$ value we adopt across the entire disk, and $\alpha_{\rm gap}$ is the peak $\alpha$ value at the center of the gap, assumed to be a factor  $\times$ $\alpha_0$. A higher $\alpha_{\rm gap}$ leads to a deeper gap. We typically consider a gap depth of 10 $\times$ $\alpha_{\rm 0}$.   

For selecting the locations of these gaps, we considered data from two detailed observational surveys of structured disks: \citep{huang18a,long18}. We combined the data from both surveys to find three common gap locations in disks, at approximately 10, 40, and 70 au. For each particle size, we then ran a set of 8 simulations including a no-gap run, single-gap runs at each radial location, two-gap runs at two of these three locations, and a three-gap run incorporating gaps at all these three radii. We explored three particle sizes (0.3, 1 and 3 mm) and therefore performed 24 simulations in all. 

\subsection{Exploring Depths of Gaps}
\begin{figure*}
	\centering
	\includegraphics[scale=.4] {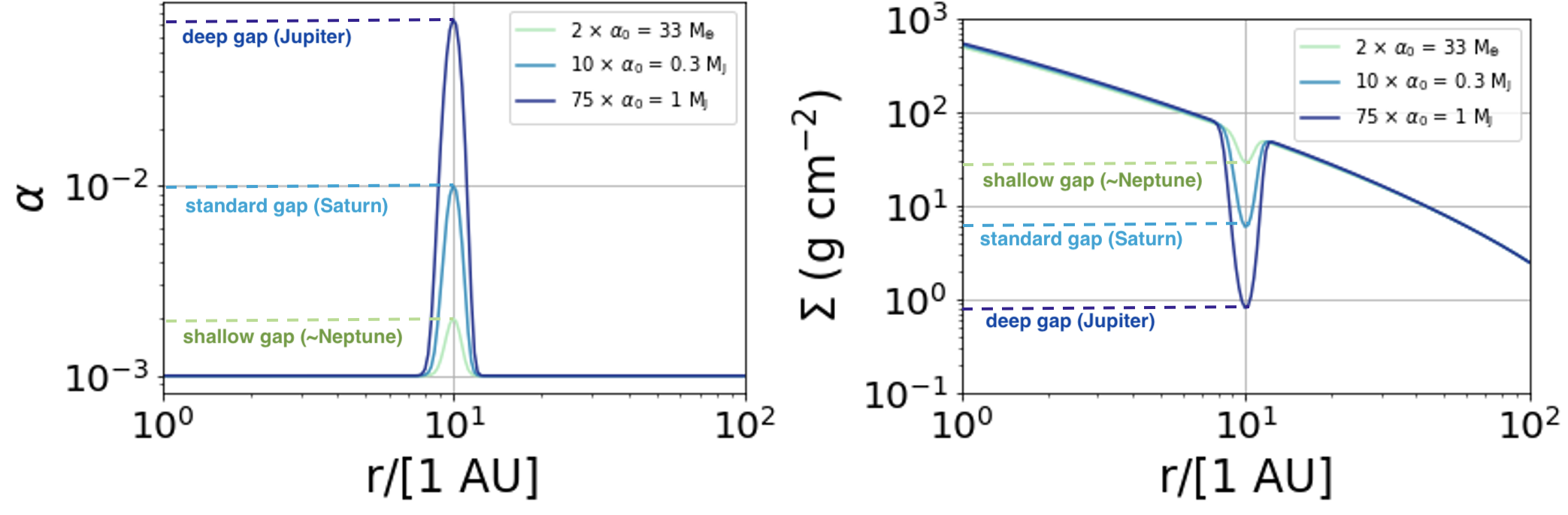}
	\caption{Radial profiles of turbulent viscosity $\alpha$ (\textit{left}) and gas surface density $\Sigma$ (\textit{right}) in different disk models that include gaps of different depths shown here for 10 au (and also applicable for other radii). Different values of $\alpha$ at the gap location produce different depletions of $\Sigma$, creating gaps of different depths, equivalent to different masses of planets within them.}
	\label{fig:2}
\end{figure*}

To test the effect of gap depth on water vapor abundance in the inner disk, we choose four gap locations - 3, 10, 40, and 70 au and for each location, we explore a range of gap depths, corresponding to peak gap $\alpha$ values equivalent to: 2 $\times$ $\alpha_0$ (shallow), 10 $\times$ $\alpha_0$ (standard), and 75 $\times$ $\alpha_0$ (deep). For this section, we choose an additional gap in the inner disk at 3 au, to test the trapping efficiency for gaps of different depths at this location, crucial for terrestrial planet formation. (See Table 1 for a summary of all these simulations performed in this study.). By adopting results from hydrodynamic simulations performed by \citet{zhang18} (see their Figure 2), these three gap depths are consistent with housing a $\sim2$ Neptune mass (33 M$_{\oplus}$, the lowest mass considered in \citet{zhang18} that in this work we will simply refer to as a $\sim$ Neptune mass), Saturn mass (0.3 M$_{\rm J}$) and a Jupiter mass planet, respectively. These three gap depths also correspond to a factor of a few, one, and two orders of magnitude respectively, in the depletion of gas surface density $\Sigma$ at the gap location (i.e., a reduction in $\Sigma$ with respect to that at locations adjacent to the gap), as shown in Figure \ref{fig:2}. 

For this investigation, we ran simulations for each gap depth at 4 radial locations for three particle sizes (0.3 mm, 1 mm, and 3 mm), running an additional 27 simulations in all. 

\section{Results of Parameter Study}\label{sec:results}
In this study, we explore in detail the effect of the presence of multiple gaps as well as the effect of the depth of gaps on the vapor enrichment in the inner disk regions. In this section, we present the results of simulations carried out for each investigation. 

\subsection{Exploring Multiple Gaps}\label{results:multgaps}

\subsubsection{Fiducial runs with 1 mm particles }

\begin{figure*}
	\centering
	\includegraphics[scale=0.42] {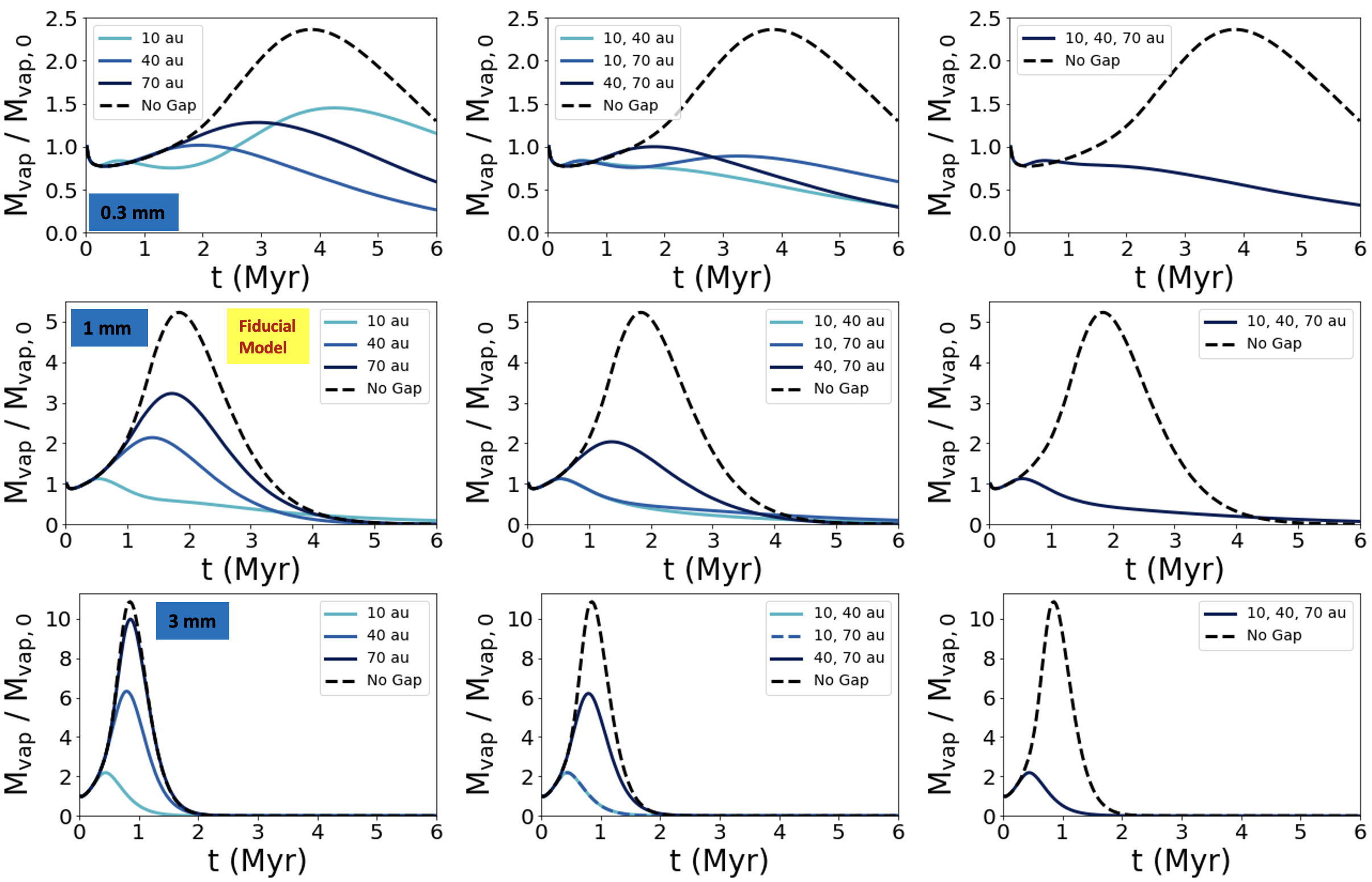}
	\caption{Plots show time evolution of vapor enrichment (mass of vapor at each time normalized to mass of vapor at time t = 0) in disk simulations with multiple gaps, performed with 0.3 mm (top row), 1 mm (middle row) and 3 mm (bottom row) particles. Left column shows simulations of disks with one gap, middle column shows simulations of disks with two gaps, and right column shows simulations of disks with three gaps. Gaps, if present, are located at 10, 40 and/or 70 au. For these simulations, we use the standard gap depth (10 $\times$ $\alpha_0$), equivalent to housing a Saturn-mass planet within the gap. Line colors denote simulations with single or multiple gaps at different radial locations. No-gap simulations are shown as dashed black lines for comparison. 
 }
	\label{fig:3}
\end{figure*}

We perform a suite of disk evolution simulations with 1 mm particles to explore the effect of the presence of multiple gaps on inner disk vapor enrichment and present these results in the second row of Figure \ref{fig:3}. To get vapor enrichment, we first integrate the surface density of water vapor across the inner disk within the snow line region, to get total vapor mass at each time, and then normalize the total vapor mass at each time t to its initial value at time $t$ = 0. As studied in detail in \citet{kalyaan21}, vapor enrichment of the inner disk increases as icy particles drift inwards and ice in them sublimates to vapor within the snow line. It reaches a peak when most of the dust population has drifted in and decreases with time due to stellar accretion.

For 1 mm particles, the time-evolving inner disk vapor enrichment in disks with only one gap are plotted in the left column and middle row of Figure \ref{fig:3}. Vapor enrichment profiles for disks with no gaps are shown in black dashed lines for comparison, and are referred to as `\textit{the no-gap limit}'. Gaps, if present, trap dust particles in the pressure bumps beyond them and block their entry into the inner disk. As previously found in \citet{kalyaan21}, for this particle size, we find that water vapor enrichment is lowest for disks with closer-in gaps (here, 10 au) as it is the closest gaps that are able to block the most mass of icy particles beyond them. Outer gaps (such as 40 au and 70 au) located further out are unable to block as much water mass and therefore show relatively higher vapor enrichment at all times. In the following sections, we will discuss that the `sweet spot' (i.e., where particles of a specific size are best trapped; \citep[][]{kalyaan21}) is around 10 au for 1 mm particles in our simulations. 

The subplot at the center (middle row, and middle column) of Figure \ref{fig:3} shows the results of our simulations with two gaps with the same particle size. We find that any combination consisting of the innermost 10 au gap (i.e., 10 and 40 au, and 10 and 70 au) results in the lowest vapor enrichment in the inner disk. In comparison, 40 and 70 au gaps combined don't trap as many icy particles behind them and result in high vapor enrichment in the inner disk. We find that the pair of gaps at 10 and 40 au is slightly better at trapping more particles than the gaps at 10 and 70 au. 

The subplot in the rightmost column and middle row in Figure \ref{fig:3} shows the results of our simulations with three gaps with the same particle size. Though this model has an additional third gap compared to previous simulations, vapor enrichment profiles are not too different from our most efficient two gap run. This is because our 70 au gap does not trap any more particles that are not already efficiently trapped by the closer-in gaps. 
The two inner deeper gaps combined would have trapped most of the material the 70 au gap would not have been able to and therefore shows similar vapor enrichment over time in the inner disk either with or without the 70 au gap.

\subsubsection{Simulations with different particle sizes} 

We also perform simulations with smaller (0.3 mm) and larger (3 mm) particle sizes and present these results in the top and bottom row of Figure  \ref{fig:3}. 

The 0.3 mm particle simulations provide some interesting results. Among the single gap simulations, it is the 40 au gap that is most efficient at trapping icy particles beyond it. The gap at 10 au is not as efficient at blocking material and is “leaky” as it allows a large amount of mass to pass through at later times. The gap located at 70 au is an efficient barrier; however, as it is located too far out, there is less mass beyond it to block. Among the models with two gaps, all pairs generally yield similar vapor enrichment. However, the combination of 10 and 40 au is not only the most efficient 2-gap combination, it is more efficient than either the 10 or the 40 au gaps, as more particles are trapped at multiple locations. For the model with three gaps, as seen previously with 1 mm particles, we see that the additional gap at 70 au has little effect and that the water vapor enrichment profile is not significantly different from the most efficient model with two gaps. 

For the 3 mm particle size, closer-in gaps are much less leaky than for smaller particles. Thus, the gap at 10 au efficiently traps the most material beyond it. On the contrary, the gap at 70 au, though also efficient, traps the least material beyond it. Larger particles drift inwards very quickly that at 0.1 Myr when the gap opens, very few particles are available to be trapped beyond 70 au in our model. 40 au gap, in comparison, is able to trap more material before they drift in by 0.1 Myr.  Among the multiple gap models, we again see that any pair of gaps that includes the 10 au gap shows the least enrichment. Since the gap at 10 au is already very efficient at blocking particles from passing through, any additional outer gap has a negligible effect on the vapor enrichment in the inner disk. 

Note that different particle sizes have different drift speeds and will therefore show different times of peak vapor enrichment in the inner disk (see \citet{kalyaan21} for more details). 

In general, we find that the presence of multiple gaps can provide the opportunity for trapping icy pebbles at more locations, as seen in the case of the model with 10 and 40 au, with 0.3 mm particles, where the presence of an additional gap led to fewer pebbles reaching the water snow line and therefore lower vapor enrichment. 

\subsection{Exploring Depths of Gaps}\label{sec:gapdepth}

We next investigate the effect of varying the depth of the gap on the vapor enrichment in the inner disk. We show these results in Figure \ref{fig:4} (see also Appendix Figure \ref{fig:8}).

\subsubsection{Fiducial Runs with 1 mm particles}
\begin{figure*}
	\centering
	\includegraphics[scale=.365] {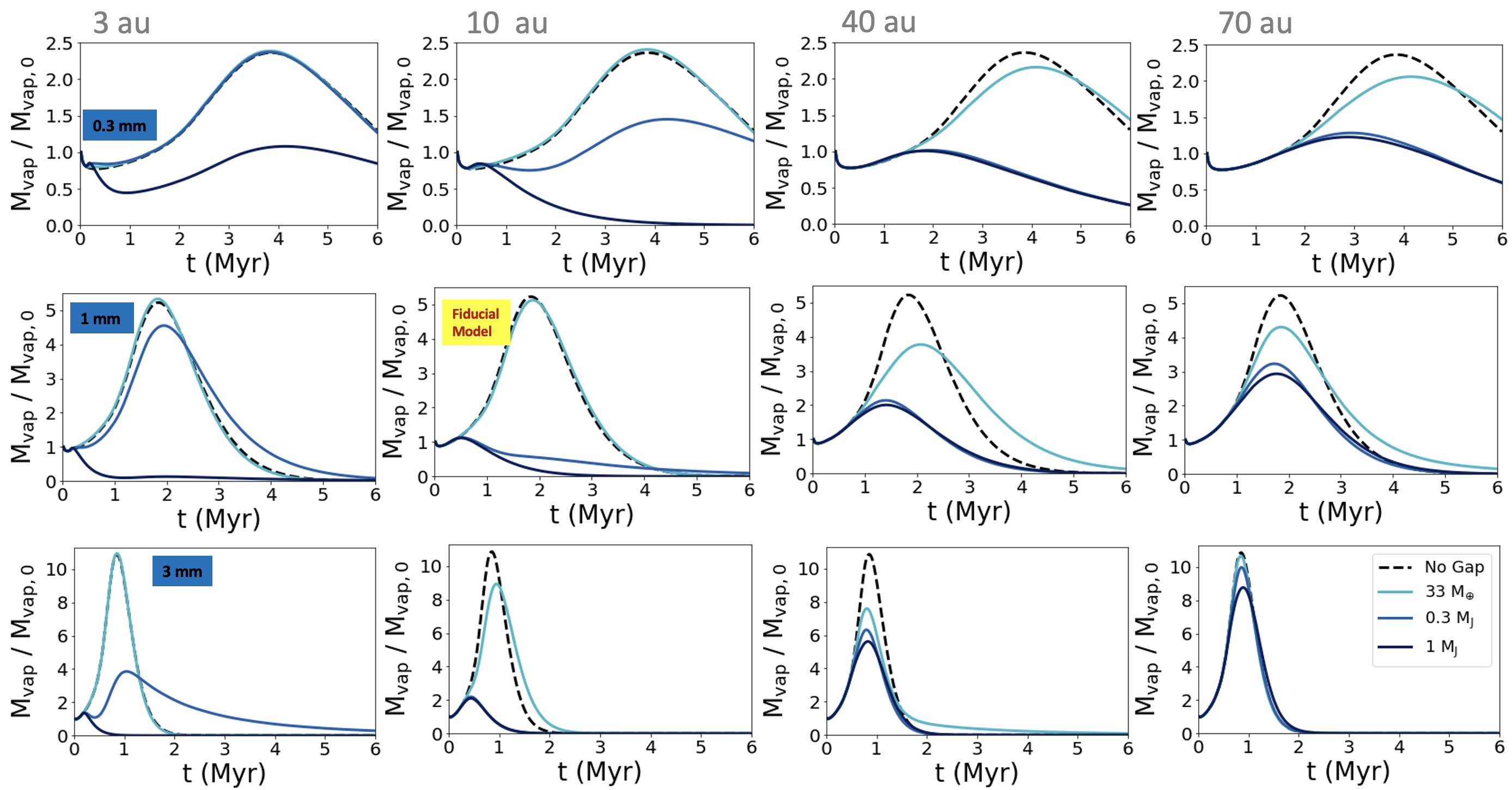}
	\caption{Plots show time evolution of vapor enrichment (mass of vapor at each time normalized mass of vapor at time t = 0) in disk simulations with gaps of different depths, performed with 0.3 mm (top row), 1 mm (middle row) and 3 mm (bottom row) particles. From left to right, columns show simulations where a single gap is placed at 3, 10, 40 and 70 au, respectively. Colors denote the mass of planet that would carve the gap with a specific depth, i.e., green for shallow gap hosting a $\sim$ Neptune mass planet, blue for standard gap hosting a Saturn-mass planet, and dark blue for deep gap hosting a Jupiter mass planet. See text for details. No-gap simulations are shown as dashed black lines for comparison.} 

	\label{fig:4}
\end{figure*}

To understand the grid of results of our simulations depicted in Figure \ref{fig:4}, we will first describe the results of our fiducial model in Figure \ref{fig:4} (middle row, second column) where we vary the gap depth for 1 mm particles at 10 au. This plot includes the enrichment profiles for a simulation with no gap, with a shallow gap (2 $\times$ $\alpha_0$), equivalent to hosting a $\sim$ Neptune-mass planet, standard gap (10 $\times$ $\alpha_0$ hosting a Saturn-mass planet; also used in our multiple gap runs) and deep gap (75 $\times$ $\alpha_0$) hosting a Jupiter-mass planet. 

We find that by decreasing its depth, the gap becomes much more leaky to inward pebble drift, so much so that the 
vapor enrichment profiles for the simulations with the shallow gap are very similar to the model with no gaps, i.e. the `\textit{no-gap limit}'. We also find that increasing the gap depth leads to a more efficient pebble trap blocking more solids beyond it. 

The middle row of Figure \ref{fig:4} shows the simulations for the remaining radial gap locations (3, 40, and 70 au) for 1 mm particles. As before, when less deep, gaps at all these locations become less efficient barriers, and allow significant solid mass to drift through the gap. For 10, 40, and 70 au gaps, we also see that for each, there is a gap depth beyond which a deeper gap has little effect on the vapor enrichment in the inner disk, i.e., increasing the depth from our fiducial value to deep does not significantly alter the vapor enrichment profile. We will henceforth call this limit the `\textit{lower enrichment limit}', where particles are already efficiently trapped beyond the gap, and the inner disk vapor enrichment is essentially dependent only on the ice mass reservoir between the star and the gap. 

In this set of simulations, we find that a 10 au gap appears to be a better barrier simply because it can trap more of the particles from drifting in (once opened at 0.1 Myr) as it is the closest gap to the star even though it is leakier than a gap further out (at 40 au; also see Appendix Figure \ref{fig:8}, middle row, columns 2 and 3). On the other hand, the 70 au gap is a very efficient barrier and is close to its lower enrichment limit. However, it traps less material beyond it because of how far out the gap is located. This is also because most particles would have already drifted inward of this radius before it opened at 0.1 Myr. The 3 au gap is extremely leaky and unable to block much pebble mass, except if it is a deep gap; then it blocks the most mass beyond it in all simulations here.

\subsubsection{Varying Particle Sizes}

The top row of Figure \ref{fig:4} shows the simulations performed with 0.3 mm particles. Here, both the 3 and 10 au gaps are significantly leaky for 0.3 mm particles. We see that increasing the gap depth does make these gaps much more efficient.
From previous simulations, we know that 40 au gap is an efficient gap location for the 0.3 mm particles, as the profiles reach the lower enrichment limit with an increase in depth over the standard gap. Decreasing the gap depth allows the vapor enrichment profile to approach the no-gap limit. The 70 au gap location results are similar to the 40 au gap; the 2 x $\alpha_0$ shallow gap is slightly better at trapping particles at 70 au. Again, the 70 au simulations reach the lower water enrichment limit with increasing gap depth.

The bottom row of Figure \ref{fig:3} shows simulations performed with 3 mm particles. The 10 au radial location is the most efficient gap location for this particle size. We again see that increasing the gap depth for this location has little effect on the water enrichment. A shallower gap (2 x $\alpha_0$) here is unable to trap many particles of this size, yielding water enrichment profiles approaching the no-gap limit. The 40 au location is a relatively efficient gap, but it does not trap behind it as much mass as the 10 au, as 3 mm particles drift much more rapidly. The 70 au location traps even fewer particles, being even further out. Increasing the depth of the gap at these two locations does trap more particles, but neither location can trap as much as beyond the closer-in radial locations. Both of these locations therefore don't have as significant an effect on influencing the vapor enrichment in the inner disk, as the closer-in radial locations.

\section{Discussion}\label{sec:discussion}
In this section, we discuss some of the main outcomes from the results of our modeling study.

\subsection{How Gap Depth Influences Vapor Enrichment}
\begin{figure*}
	\centering
	\includegraphics[scale=.42] {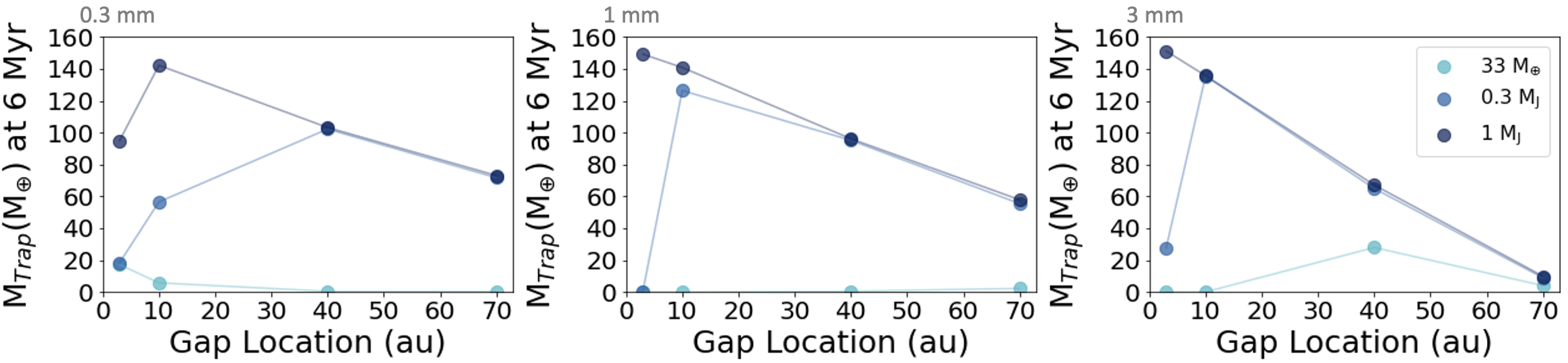}
	\caption{Plots show final trapped mass beyond the gap at 6 Myr for simulations with 0.3 mm (left), with 1 mm (middle) and 3 mm (right) particles, respectively. Final trapped mass shows how effective each gap (with certain depth and location) is at trapping particles of various sizes. Green points denote shallow gaps (hosting $\sim$ Neptune-mass planets), blue points denote standard gaps hosting Saturn-mass planets, and dark blue points denote deep gaps hosting Jupiter-mass planets.  }
	\label{fig:5}
\end{figure*}

The inner disk water vapor enrichment is affected both by the gap depth and location. Figure \ref{fig:5} shows our grid of simulations in Figure \ref{fig:4} in terms of mass trapped beyond the gap at the end of the simulation at 6 Myr. We use this metric to understand how effective any gap is at trapping particles and retaining them beyond it over time. We show how the trapped mass varies with time for all these simulations in the Appendix (Figure \ref{fig:8}). 

\citet{kalyaan21} had found that depending on particle size, there were specific radial zones which they called `sweet spots' where a gap of the same depth would trap \textit{more} particles of that size as well as \textit{more efficiently} compared to other locations. They found that if a gap was present too close-in, it could be more leakier and allow for some passage of particles; present too far, a lower mass of particles will be available to be trapped beyond the gap. This can be seen in Figure \ref{fig:5} as the gap location with the largest mass trapped beyond it for a given particle size and gap depth. For 1 mm particles, we find that the 10 au gap (as well as the 40 au gap to an extent) appear to be the most efficient barriers, trapping the most mass beyond them, consistent with \citet{kalyaan21} who had found that gaps at $\sim$ 15 - 30 au were efficient at blocking similar-sized particles. We also see that the `sweet spot' moves inward as particle size increases (another finding made by \citet{kalyaan21}) as larger particles are better trapped behind closer-in gaps, for a given gap depth. For example, we see that the `sweet spot' moves from 40 au for 0.3 mm particles, closer to 10 au for 1 and 3 mm particles. Additionally, in our work, we find that the `sweet spot' gets closer in with increasing gap depth, for a given particle size. This effect is most apparent for 1 and 3 mm particles. For 1 mm particles, the `sweet spot' moves from 70 au to 3 au, as gap depth is increased from shallow to deep. For 3 mm particles, it moves from 40 au to 3 au from shallow to deep.

Overall, we find that for a given particle size, different gap depths are required at different gap locations to be effective at trapping particles and to trap the most mass beyond it. Deeper gaps are required in inner regions of disks to trap smaller particles, but large particles are trapped easily beyond them. Particles of all sizes are trapped easily beyond moderately deep outer disk gaps; however, gaps have to open quickly before large particles drift inwards out of that region. 

We also find that increasing the depth of gaps make them more efficient pebble traps, depending upon the gap location and the particle size. We also find that lowering the gap depth make gaps at all locations leakier and inefficient pebble traps, which show vapor enrichment profiles comparable to that of uniform disks (see Appendix A for more details).
 
We note that we have not explored different disk sizes, which are likely to impact inner disk vapor enrichment as found by \citet{BitschMah2023}.
 
\subsection{Vapor Enrichment in Disks with Forming Planets}\label{plansystems}

\begin{figure}
	\centering
	\includegraphics[scale=.395] {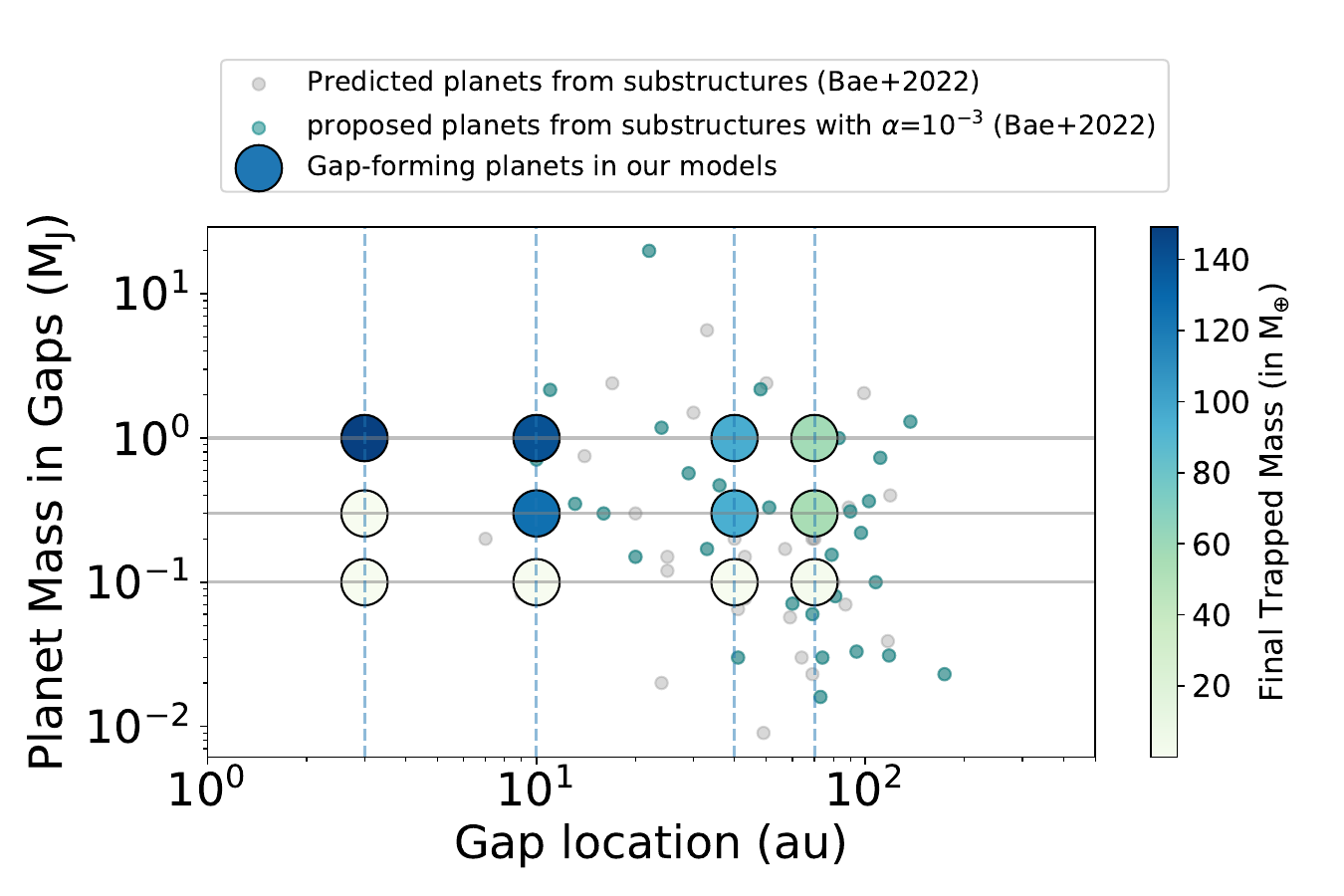}
	\caption{Comparison of gap properties used in our simulations with proposed planets from observed substructure in disks. We show all data compiled in Figure 7 of \citet{bae2023review} as small grey circles and highlight a subset of them which correspond to a specific turbulent viscosity $\alpha$ = 1 $\times$ 10$^{-3}$ as small teal circles. Large circles and the corresponding color bar show the masses of trapped icy particles beyond the gap in 12 different simulations at 6 Myr, derived from the middle row of Figure \ref{fig:8} with 1 mm particles.}
	\label{fig:6}
\end{figure}

Finally, to understand our results from the broader context of planetary systems and planetary system architecture, we consider the masses of young planets that may be resident within disk gaps. We compare with the hydrodynamical simulations of \citet{zhang18} to determine the masses of planets that could create similar depletions of gas surface density at the location of the gap as we have in our models. As mentioned before, we consider three different gap depths that can be produced by three different planets: i) a shallow gap carved by a $\sim$ 33 M$_{\oplus}$ planet, or $\sim$ Neptune; ii) our standard gap carved by a 0.3 M$_{\rm J}$ planet, or a Saturn; and iii) a deep gap carved by a 1 M$_{\rm J}$ planet, or a Jupiter. We first provide context to our study by comparing with observed structure in disks, and then consider icy pebble delivery in disks with different planetary configurations. 

\subsubsection{Comparison with Planets inferred from Disk observations}\label{planetdisksims}

We compare with structure (i.e., gaps) observed so far in disks in Figure \ref{fig:6}. From 1 mm particle simulations shown in Figure \ref{fig:4}, we compute the final masses of icy particles trapped beyond gaps in disks composed of a single gap of various depths, located at different radial locations, (see also Figure 8) and present them within a plot of the mass of the planet carving the gap in the disk gas against radial location. Also plotted are the masses of proposed planets against their observed locations from \citet{bae2023review}, which includes the results from various hydrodynamical studies that predict planet masses from ALMA observations. In effect, these data points serve as a summary of gap properties (location and gap depth) derived from observations to provide a comparison with our theoretical results. Since the derived planet mass is highly sensitive to the turbulent viscosity used in the simulations, we highlight a subset of the data points from \citet{bae2023review} that specifically used $\alpha$=1 $\times$ 10$^{-3}$ in Figure \ref{fig:6}, which is consistent with simulations in this work.

Comparing with our simulations (and making the assumption that each gap had a dominant effect on vapor enrichment, over other accompanying gaps in each disk), we argue that the majority of substructures observed by ALMA so far might only trap a small pebble mass beyond them; these disks may therefore have inner regions that are still water rich. This is because water enrichment is strongly dependent on the location of the gaps and our models predict the innermost gaps would have the most significant impact in blocking icy pebble delivery to the snowline region.
However, ALMA observations have provided only a few close-in gaps that may host Jupiter-sized planets or larger at distances of $\sim$ 10 au, and none yet very close to the star (within $\sim$3 au) due to insufficient angular resolution (with the exception of the 1 au cavity in the TW Hya disk, which is resolved because it is only at 60~pc \citep[][]{Andrew16}, and possibly a few other gaps at $\sim5$~au inferred from super-resolution techniques \citep{jennings22_taurus}). 

We note that we do not take into account the accretion onto forming planets, which may also change the gap depth depending on disk viscosity as found by \citet{BergezCasa2020}. We also note that, in addition to the depth of gaps, their width may also matter in making them effective traps of pebbles. Wider gaps have not been tested by this paper, but studies have proposed that they may house more than one planet within them \citep[][]{zhang18}. Future high resolution observations from ALMA with better resolved close-in gaps would be helpful to test our predictions on a broader sample.

\begin{figure*}
	\centering
	\includegraphics[scale=.32] {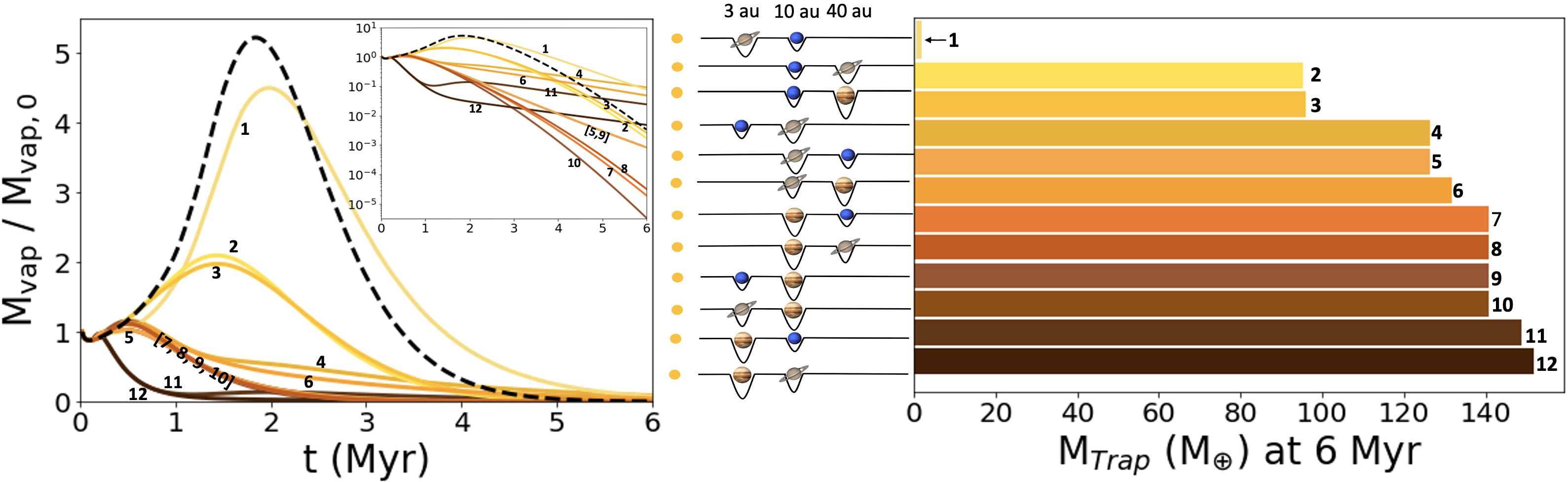}
	\caption{Inner disk vapor enrichment and final pebble mass trapped in disks with different young planets. Left panel shows inner disk vapor enrichment as a function of time for disks with different young planetary configurations, where inset plot shows the same vapor enrichment profiles in log scale for clarity. Right panel shows the final mass of icy particles trapped at 6 Myr for each configuration. Configurations are sorted in order of increasing trapped mass and labeled 1 through 12 (top to bottom). Configuration 1 with shallow gaps created by Saturn and Neptune-mass planets trap very little mass with no visible horizontal bar on the plot.
 }
	\label{fig:7}
\end{figure*}

\subsubsection{Vapor Enrichment in Disks with Young Planets}

Next, we test different planetary configurations with multiple planets located at different radial locations - 3, 10, and 40 au. We choose these radial locations as our simulations reveal that closer-in gaps have much greater effect on vapor enrichment than those in farther out regions (e.g., 70 au), more so for shallower gaps at each location. Using simulations with 1 mm sized particles, we then study how the water vapor content of the inner disk from pebble delivery changes with time, and how much combined icy pebble mass is trapped beyond all of the gaps at the end of the simulation. The results of the suite of 12 simulations are shown in Figure \ref{fig:7}. 

Previously, our single-planet simulations with 1 mm particles (Figures \ref{fig:4} and \ref{fig:5}) have revealed that a young $\sim$ Neptune mass planet is unable to retain any significant mass beyond it at all distances. A Saturn-mass planet can retain appreciable mass at all radii except very close-in (3 au and within), whereas Jupiter is able to efficiently retain mass beyond it even as close as 3 au.

In our suite of multiple-planet simulations (Figure \ref{fig:7}), configurations 1 and 2 explore the regime of shallow gaps. Configuration 1 has two shallow gaps; an inefficient leaky gap carved by a close-in Saturn-mass planet at 3 au and a shallow gap created by $\sim$ Neptune-mass planet at 10 au, which together trap negligible mass beyond them. Configuration 2 having an efficient gap at 40 au formed by the Saturn-mass planet is able to trap considerable mass ($\sim$ 100 M$_{\oplus}$) beyond 40 au. In configurations 2-12, the location of the most massive planet is critical. In configurations 2, 4, and 5, the location of the Saturn-sized planet has the most impact on pebble and volatile delivery into the inner disk (in 4 and 5, the change in the location of Neptune has no real effect). In configurations 3, and 6-12, it is the location of Jupiter that matters most. A deep gap (hosting a Jupiter- or similar-sized planet) is an effective barrier irrespective of location and therefore efficiently blocks almost all of the pebble mass beyond it from entering the inner disk. If Jupiter is present at 40 au, it blocks all pebbles beyond 40 au. An additional Neptune at 10 au (configuration 3) makes for a leaky barrier, which may partially block some of the material between Neptune and Jupiter; an additional Saturn at 10 au (6) makes for a better barrier, trapping more of this intervening ice mass, therefore showing lower vapor enrichment in the inner disk and trapped ice mass beyond the gaps. 

Configurations 7-10 all show similar amounts of trapped mass, as before, because the presence of Jupiter dominates over any additional presence of closer-in inefficient (leaky) gaps (9 and 10). An additional further-out planet has a negligible effect on the already efficient barrier that the inward Jupiter present (7 and 8). Configurations 11 and 12 show a closer-in Jupiter at 3 au. At this location, though still an effective barrier, it becomes slightly leaky and partially allows the passage of a few particles (also see Figure \ref{fig:8}) showing the efficiency of each gap as a barrier for each particle size until 6 Myr). Therefore when paired with another efficient barrier (Saturn at 10 au), configuration 12 yields the highest trapped mass, and the lowest vapor enrichment. On the contrary, as seen previously, Neptune at the same location has a negligible effect with a leaky gap at 10 au for 1 mm particles (Figure \ref{fig:4}, and Appendix Figure \ref{fig:8}). 

Overall, we find that in any configuration the closer-in the most massive proto-planet is, the more mass is trapped beyond the gap, leading to the lowest vapor enrichment in the inner disk. If the massive planet is further out, it is able to trap less material beyond it and therefore has a lower impact on the water enrichment in the inner disk from icy pebble delivery. In such a case, even a slightly leaky gap in the inner disk may be better than a deeper gap in the outer disk (see also Figure \ref{fig:8}). Additionally, disks with more gaps or forming planets have more locations for pebbles to be trapped and therefore may show lower inner disk vapor enrichment \citep{kalyaan23}.

In general, the results obtained with our model (where we introduce a gap via the viscosity profile - as done in several previous works mentioned before - as well as include dust diffusion) are similar to those obtained by the 2D  hydrodynamical simulations by \citet{weber2018}, where they find that particles with high St are filtered beyond the gap, while particles with lower St that are well-coupled to the gas partially filter through (see Appendix A, and Figure 8 for more details.)

\subsubsection{Implications for the early Solar Nebula}
Planetary system configuration 12 is somewhat similar to an earlier phase in our own solar system, where Jupiter may have formed at around 3-5 au, and Saturn may have formed at around 7-8 au within the gaseous disk \citep[e.g.,][review]{raymondmorbi2022}. We note that our comparison with solar system planets have a few important caveats. Because we employ static unchanging gaps that instantaneously form at 0.1 Myr, we assume that these planets all form at a particular time, then quickly grow to a certain mass, carving gaps in the disk of particular depth (as in \citet{zhang18}) and that they do not migrate over millions of years. However, we generally argue that massive closer-in proto-planets present strong barriers to pebble delivery by trapping large masses of icy material beyond these planets. As a result, inner regions in these disks likely receive low pebble-mass fluxes enabling no more than the formation of small terrestrial planets \citep[][]{lamb2019}. They also potentially influence the water abundance in these small planets. Our own solar system is composed of smaller and water-poor terrestrial planets within 5 au, in comparison to the massive gas and ice giants outside of 5 au, consistent with this scenario. Several previous works have explored the impact of a giant planet on the volatile and pebble delivery into the inner disk \citep[e.g.,][]{morbi2016}. Meteoritic evidence also shows that the formation of Jupiter likely produced a sufficiently deep gap that interfered with material transport between the outer and inner nebula well within 1 Myr \citep[][]{kruijer2017}. 

We also note that there are a few studies that argue that proto-Jupiter's gap may not have been the only mechanism preventing transport of particles between the inner and outer disks, as grain fragmentation at the pressure bump outside of the gap would have led to continuous leakage of small particles that would be able to pass into the inner disk \citep[][]{stammler2023}, and that if the gaps of both proto-Jupiter and proto-Saturn overlapped, it could have led to more material transport into the inner disk compared to only the presence of Jupiter. This is argued to occur as the pressure gradient may be made less steep by the addition of the Saturnian gap \citep[][]{haugbolle2019}. We have not considered particle fragmentation or overlapping gaps in this work.

\subsection{The Effect of Multiple Gaps on Vapor Enrichment}

In this work, we performed simulations where we tested the effect of additional gaps of similar depth on inner disk vapor enrichment (Section \ref{results:multgaps}). We subsequently also performed a set of 12 simulations of disks with gaps of different depths, corresponding to different planetary configurations with 1 mm particles (Section \ref{plansystems}). We found that if a disk contained a gap located in the sweet spot region, adding more outer gaps did not significantly lower the water vapor enrichment. However, an inefficient gap when paired with an efficient outer gap resulted in more particles trapped beyond gaps overall, resulting in a significant further depletion in vapor content in the inner disk. This can be seen in Figure \ref{fig:3} in the model with two gaps at 10 and 40 au with 0.3 mm particles, as well as in Figure \ref{fig:7} by (i) comparing configurations 1,2 and 3, (ii) configurations 5 and 6, and (iii) comparing configurations 11 and 12). In all of these combinations, multiple gaps in disks provide more locations for trapping pebbles and retaining them in the outer disk. 
Amongst all combinations, pairs of the deepest inner-most gaps presented the best barriers for icy pebble delivery, substantially decreasing vapor enrichment in the inner disk. 

\section{Summary \& Conclusions}\label{sec:conclusions}

Based on recent work suggesting that disk gaps may regulate the inner disk water vapor enrichment by trapping a fraction of icy pebble mass beyond them, we performed a grid of simulations to test the effects of multiple gaps presumed to be carved from forming planets at different radial locations and with different depths. This study expands the parameter exploration carried out in our previous work \citep[][]{kalyaan21}, and our new main conclusions are as follows:

\begin{enumerate}
    \item Disk gaps produced by \textit{close-in massive planets} trap a larger mass of icy pebbles beyond them and therefore they likely have the greatest impact on limiting water vapor enrichment from ice sublimation within the snowline. The presence of such massive planets may result in water-poor terrestrial planets, as predicted by other studies.

    \item Additionally, disks with multiple gaps, presumably with several forming planets, provide more locations for pebbles traps and are therefore also likely to have less water vapor enrichment in the inner disk, again leading to the formation of water-poor terrestrial planets. This effect is more significant if one of the gaps is carved by a giant planet with a smaller disk radii. 

    \item We also find that a moderately large proto-planet in a somewhat leaky closer-in gap may trap a higher mass of pebbles and therefore substantially reduce the inner disk vapor enrichment, than a deeper gap formed by a farther-out massive planet.
    
    \item Our results are consistent with the overall scenario that while the outermost gap sets the outer radius of the dust/pebble component in the disk, the innermost gap regulates the inner disk water vapor enrichment depending on its efficiency in trapping icy pebbles outside the snowline.
 
\end{enumerate}

\begin{table*}\label{table:1}
\centering
\begin{tabular}{ |p{4.2cm}|p{2.8cm}|p{2.8cm}|p{2.8cm}|p{2.8cm}|  }
 \hline
 \bf{Model Parameter} & \bf{Standard Model with no gaps} & \bf{Simulations with Multiple Gaps} & \bf{Simulations with Different Gap Depths} & \bf{Simulations with Young Planets}\\
 \hline
 
 Particle Size & 0.3, 1, 3 mm & 0.3, 1, 3 mm & 0.3, 1, 3 mm & 1 mm \\
 Gap Depth &  & Standard & Shallow, Standard, Deep & Shallow, Standard, Deep \\
 Gap Location &  & 10, 40, 70 au & 3, 10, 40, 70 au & 3, 10, 40 au \\
 Number of Gaps & 0 & 1, 2, 3 & 1 & 2 \\
 \hline
 Total Number of Simulations & 3 & 7 + 7 + 7 = 21 & 12 + 12 + 12 = 36 & 12 \\ [1ex]
 \hline
 \hline
  \textit{Comment} & For each particle size, models are run with no gaps in the disk. 
  & For each particle size, models are run where one, two, or three gaps are introduced in the disk at radial locations 10, 40, and/or 70 au, with standard gap depth.
  & For each particle size, models are run where only one gap is introduced at a time at different radial locations, with different gap depths.
  & For 1 mm particles, models are run where two gaps are introduced at radial locations 3, 10, and/or 40 au, with specific gap depths, associated with the presence of planets assumed to reside within the gap.\\
 \hline 
\end{tabular}
\caption{Summary of simulations performed in this work}
\end{table*}

\acknowledgments

We are thankful to the anonymous reviewer for suggestions that improved the manuscript. This research work was partially supported by funding for W.E. by the STEM Undergraduate Research Experience Program at Texas State University. A.K. and A.B. acknowledge support from the NASA/Space Telescope Science Institute grant no. JWST-GO-01640.

\appendix

\section{Trapping Efficiency of Gaps of Various Depths}

\begin{figure*}
	\centering
	\includegraphics[scale=.34] {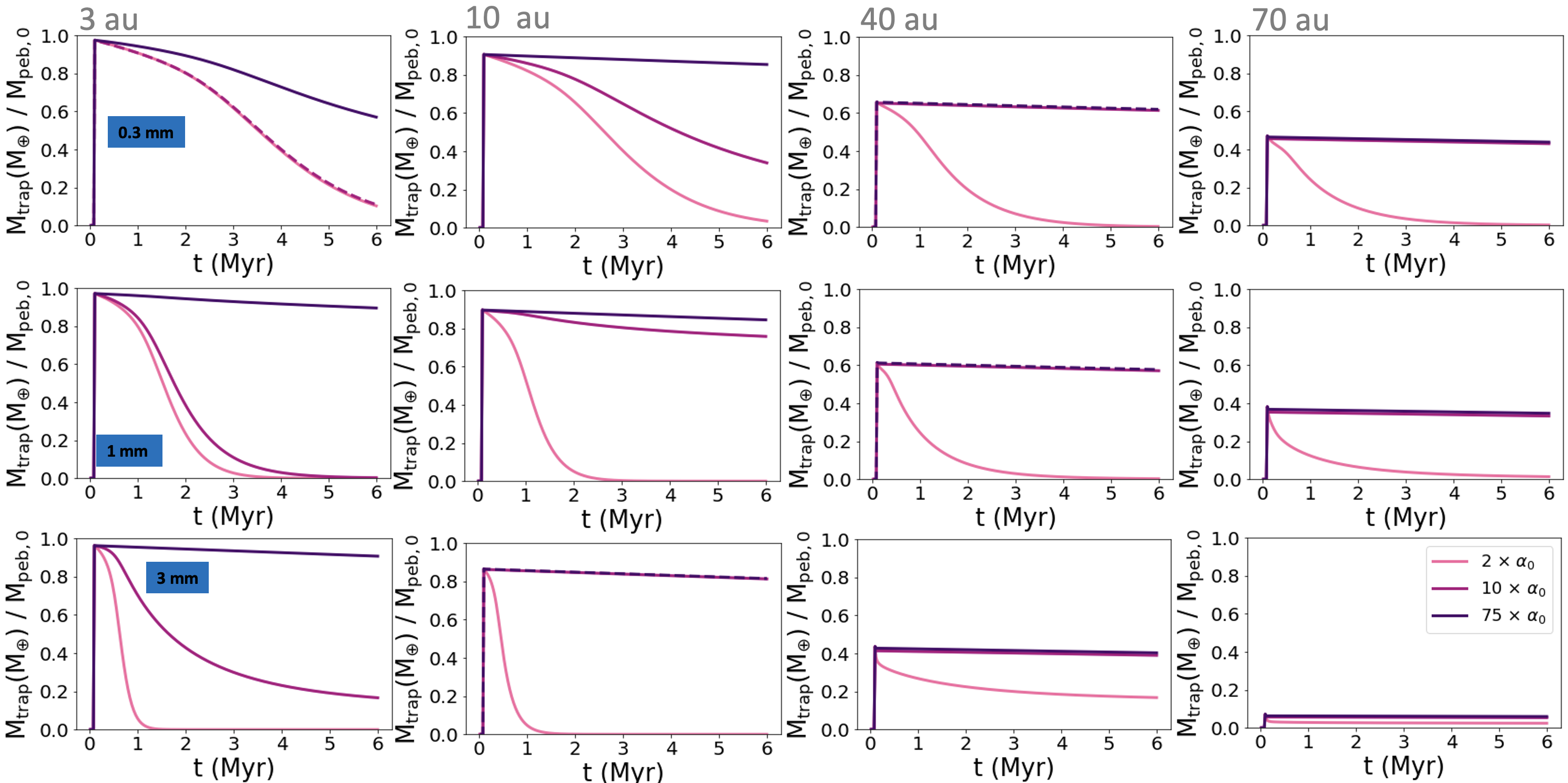}
	\caption{Grid of plots show masses of trapped particles beyond gaps of various depths over time, at different radial locations. Mass trapped is normalized to total pebble mass in the disk at time t=0, equivalent to 166.5 M$_{\oplus}$. Shallower gaps allow some material to pass through with time, while deeper gaps retain mass beyond them even after 6 Myr. Each plot row denotes simulations for different particle sizes - 0.3, 1, and 3 mm, from top to bottom, respectively. Columns denote radial location of the gap. }
	\label{fig:8}
\end{figure*}

Figure \ref{fig:8} shows total pebble mass that is trapped beyond a gap over 6 Myr in all the simulations described in Section \ref{sec:gapdepth}. The total pebble mass here is normalized to total initial pebble mass at time t = 0, equivalent to 166.5 M$_{\oplus}$ . Final trapped masses (i.e., masses trapped beyond the gap in each model) shown in Figures \ref{fig:5} and \ref{fig:6} in the text are derived from this figure. How efficient gaps are as regions of pebble trapping are dependent on the Stokes number of the particles as well as the pressure gradient around the region with the gap. Larger particles above a certain size are trapped, while smaller particles pass through. This is seen in studies such as \citet[][see equations 10, and 11 in their work]{pinilla12} that show the dependence of the smallest trapped particle size on gas surface density and the pressure gradient. 

While comparing simulations in each column together that have gaps at the same location, shallow gaps (pink lines) partially allow particles to pass through into the inner disk, while deep gaps (violet lines) trap particles more efficiently and reliably retain this mass beyond them for several million years. Farther-out deep gaps are most efficient in this regard. Additionally, particle size (simulations in each row) sets the drift speed; smaller particles drift slowly and are able to be trapped even beyond 70 au after gap opens at 0.1 Myr, while 3 mm particles would drift too rapidly into the inner disk before gap opens, to be trapped beyond it. 

\bibliography{sample63}{}
\bibliographystyle{aasjournal}

\end{document}